\documentclass[12pt]{article}

\title{\large Simulations of experimental conductance spectra of $YBa_2Cu_3O_y$
junctions: Role of a long d-wave decay length and a small $is$
component in the pair potential near the interface}

\author{I. Lubimova and G. Koren\\
 Physics Department, Technion - Israel Institute of Technology\\
 Haifa, 32000, ISRAEL}

\date{ }
\def\bfig {\begin{figure}[tbhp] \centering}
\def\efig {\end{figure}}

\usepackage{amssymb}
\usepackage[pdftex]{graphicx}

\begin{document}
\DeclareGraphicsExtensions{.jpg,.mps,.eps,.pdf}

\normalsize \baselineskip=5mm \sf \maketitle \vspace{25mm}


Extended BTK-type theoretical methods were used for the
calculation of conductance spectra of NS  and NIS junctions, where
S is either a d-wave or a d+is-wave superconductor, and N is a
normal metal. We found that the length scale over which the bulk
d-wave order parameter varies spatially near the interface (the
"decay length") has a significant effect on the main features of
the conductance spectra. A relatively large decay length of the
order of 10-20$\xi_{0}$ together with a small $is$-wave component,
can explain many peculiarities observed in the experimental
conductance spectra of various junctions. We attribute the long
decay length to either a reversed proximity effect in the high
transparency junctions, or to the presence of a low $T_c\leq 10$K
s-wave component with a much longer $\xi_{0s}\approx 10\xi_{0d}$
near the interface.

\newpage

\maketitle{\large   INTRODUCTION}

Many investigations of the special properties of d-wave
superconductors and their influence on the conductance spectra
have been reported in recent years.\cite{Tanaka,Tsuei} A large
number of them deal with NIS or SIS junctions in the tunneling
limit where the effective BTK barrier strength
\begin{equation}\label{1}
Z=\frac{mH}{\hbar^{2}k_{F}}
\end{equation}
is larger than $\sim$1 (strong barrier).\cite{BTK} Here $H$ is the
amplitude of the $\delta$-function barrier at the interface. From
a theoretical point of view, the tunneling limit is simpler to
treat because in this limit of weak coupling between the
electrodes, one can consider the electrodes as almost independent
entities. More specifically, one can assume that the spatial
changes of the superconducting order parameter, arising from the
presence of a surface, occur only in a very small distance from
the interface inside the superconductor. Thus, to a first order
approximation these changes are often ignored. Conductance spectra
of NIS junctions where S is a d-wave superconductor, are
characterized by a zero bias conductance peak (ZBCP) in the
node-direction, and by a  gap in the anti-node (main axis)
direction. Theoretical calculations of low transparency junctions
reproduce these features quite well, also in the simpler case when
a spatial independent pair potential is taken into
account.\cite{K_T_BR} The ZBCP is a manifestation of zero-energy
bound states of quasi-particles near the interface. The appearance
of these bound states along the node direction of a d-wave
superconductor results from the sign change of the energy gap
$\Delta(\varphi_{+})=-\Delta(\varphi_{-})$, where $\varphi_{\pm}$
are the angles between the wave-vector of the incoming or
scattered quasi-particle and
the normal to the interface, and $\varphi_{+}=\pi-\varphi_{-}$.\\

Theoretical calculations of the conductance spectra of junctions
with a weak-barrier are more problematic. The problem arises from
the strong mutual dependence of the pair potential on both sides
of the NS interface. This is in contrast to the strong barrier
case, where the two electrodes are almost decoupled. Thus, in
order to obtain an explicit form of the spatial dependence of the
pair potential (SDPP), a complete knowledge of the quasi-classical
Green function is necessary, not only in the S side but also in
the N side of the junction. Therefore, a proximity effect must be
directly included in the calculations. There is no consensus on
the peculiarities  of the proximity effect in junctions with
anisotropic superconductors at the present time. The penetration
length of the order parameter into the N side and its symmetry
there, depend sensitively on the specific properties of the normal
metal.\cite{Ohashi,Sonier,Decca,Polturak,Pannetier} Recently, it
was found, that even ferromagnetic or antiferromagnetic metals in
NS junctions have a long-range proximity effect  under certain
conditions,\cite{Demler,Bergeret,Kadigrobov} and this is in
contrast to the predictions of certain theories.\cite{XXX} For the
123-family of superconductors, the length scale of the proximity
effect in the N side depends significantly on the oxygen-doping
level of the S side. This results from the fact that in underdoped
YBCO $T_{c}$ is reduced compared to optimal doping, and the value
of the coherence length is increased. A-priori, the proximity
interaction between the two electrodes should depend strongly on
the barrier strength. But as was shown
previously,\cite{Golubov,Wees,Imry} the effect of the barrier is
not always so evident, and even a barrier of finite strength is
not always effective in reducing the conductance of the junctions.
In particular, in NS junctions with similar materials and similar
density of states (DOS) on both sides, the influence of a barrier
is suppressed significantly. This occurs when both electrodes are
made of differently doped cuprate compounds. According to
Deutscher and De Gennes,\cite{deGen} in such cases the amplitude
of the order parameter in the normal side near the interface can
reach a value close to that in the superconductor, and there is
almost no discontinuity or jump at the interface. Therefore, in
this type of junctions the diffusion of pairs from the
superconductor into the metal is almost unaffected by the barrier,
and the decay distance
of pairs in the normal phase may be anomalously large.\\

YBCO  has a short coherence length  $\xi_{0}=\hbar
v_{F}/\Delta_{0}$ of about $ \rm 10-30\AA$ where $\Delta_{0}$ is
the bulk value of the d-wave pair potential.\cite{Segawa} In spite
of this fact, it follows from the above discussion that in studies
of NS-junctions with similar cuprates one should examine a wide
range of decay lengths in S and proximity lengths in N. To clarify
this different length scales dependence, we decided to calculate
the corresponding conductance spectra in a phenomenological way.
We didn't develop a new model for self-consistent calculations of
the pair potential in proximity systems. Instead, we assumed
reasonable pair potential shapes (SDPP), and calculated the
resulting conductance spectra. From the results we were able to
identify which of the peculiarities of the SDPP had the most
significant effect on the conductance spectra, and what form of
SDPP leads to the best fit
of the experimental data.\\

\maketitle{\large RESULT AND DISCUSSION}

We have calculated numerically the conductance spectra of a normal
metal (N) - d-wave superconductor (S) junction as a function of
applied voltage using the basic model developed by Tanaka and
Kashiwaya.\cite{Tanaka} According to their theory, the conductance
is given by:

\begin{equation}
  \sigma(eV)=  \frac{\int\limits_{-\pi/2}^{\pi/2}\int\limits_{-\infty}^{\infty}\sigma_{S}(E,\varphi)
  (-\frac{\partial f(E+eV)}
  {\partial E})\sigma_{N}(\varphi)\cos\varphi d\varphi dE}
  {\int\limits_{-\pi/2}^{\pi/2}\sigma_{N}(\varphi)\cos\varphi d\varphi},
\end{equation}

\noindent where the integrations are over the angles between the
trajectory of the incoming quasi-particle and the normal to the
interface of the junction, and the energy. The normal metal
conductance $\sigma_{N}$, characterizes the transparency of the
junction  and is given by:

\begin{equation}
  \sigma_{N}=\frac{4\cos^{2}\varphi}{4\cos^{2}\varphi+Z^{2}}.
\end{equation}

\noindent where Z is the effective barrier strength as given by
Eq. (1). $f(E)$ in Eq. (2) is the Fermi distribution function at
temperature T. The function $\sigma_{S}$ in Eq. (2) represents an
extension of the standard BTK expression for the conductance
\cite{BTK}
\begin{equation}
  \sigma_{S}=1-|b|^{2}+|a|^{2},
\end{equation}

\noindent to the case of spatial dependent order parameter. In Eq.
(4), $b$ and  $a$ are the coefficients of the normal and Andreev
reflections, respectively.   In the simple BTK model, the
reflection coefficients are expressed in terms of the functions
$\Gamma_{\pm}(E,\varphi)$ in the following way:

\begin{equation}
 \frac{\sigma_{S}}{\sigma_{N}}=\frac{1+\sigma_{N}|\Gamma_{+}|^{2}+(\sigma_{N}-1)|\Gamma_{+}\Gamma_{-}|^{2}}
 {|1+(\sigma_{N}-1)\Gamma_{+}\Gamma_{-}|^{2}}.
\end{equation}

\noindent The  $\Gamma_{\pm}$ in this equation are functions of
the energy $E$ and angle $\varphi$ only, since the pair potential
is assumed to be spatially constant. In the present study however,
we use the extended BTK model which takes into account the spatial
dependence of the superconducting pair potential. Following
Kashiwaya and Tanaka,\cite{Tanaka} we calculated the
$\Gamma_{\pm}$ values at the interface by solving a set of Riccati
type equations (Eqs. (3.50 - 3.52) in Ref.\cite{Tanaka}) in which
the spatial dependent pair potentials $\Delta_{\pm}(\varphi,x)$
appear as coefficients ($x$ is the distance from the interface in S).\\

In our simulations we have chosen the following functions to
represent the spatial dependence of the pair potential of the
dominant $d_{x^{2}-y^{2}}$ and sub-dominant $s$ components of the
order parameter:

\begin{equation}
 \frac{\Delta_{d_{x^{2}-y^{2}}}(\zeta)}{\Delta_{0}}=\tanh((\zeta+k_{1})*k_{2}),
\end{equation}
and
\begin{equation}
\frac{\Delta_{s}(\zeta)}{\Delta_{0}}=k_{3}\exp(-k_{4}*\zeta^{2}),
\end{equation}

\noindent where $\zeta = x/\xi_{0}$ is a dimensionless length (see
Fig.1). The superconductor occupies  $x\geq 0$ and the interface
is exactly at x=0. Here the $k_{i}$ ($i=1..4$) are dimensionless
parameters which depend on the junction orientation relative to
the crystalline axis.  The modelling functions in Eqs. (6) and
(7), satisfy the main boundary conditions  on both component of
the pair potential
($\Delta_{d_{x^{2}-y^{2}}}\rightarrow\Delta_{0}$ and $\Delta_{s}
\rightarrow 0$ as $x\rightarrow\infty$), and are consistent with
the self consistent model results.\cite{Tanuma, Matsumoto}
Moreover, we note that the fine details of the shape of the SDPP
is not essential, as it does not affect the resulting spectra in
any major way. In order to confirm this statement, we examined
different possible functions by replacing, for instance,
$\tanh(\zeta)$ with $1-\exp(-\zeta)$. Results of these
calculations show that such changes lead only to  minor changes in
the energies of the high-order bound states in the conductance
spectra. We found out that the most important parameters of the
SDPP shape which affect the conductance spectra, are the maximal
amplitude of the sub-dominant component (the $k_{3}$ coefficient
in Eq. (7)) and the quasi-particle decay length (QPDL), which
represents the characteristic length scale over which the pair
potential varies spatially  near the interface (governed by
$k_{1},k_{2}$ and $k_{4}$ in Eqs. (6) and (7)).\\

In all our calculations we have taken into account the thermal
smearing factor which is due to the finite (low) temperature, and
the finite life time broadening of the quasi-particles at the
Fermi surface. It is a well known fact from the theory of Fermi
liquids,\cite{LL} that the inverse lifetime of the quasi-particles
is proportional to $E^{2}$. In real superconductors, there are
some quasi-particle relaxation processes which include
electron-electron and electron-phonon interactions, which modify
this simple square low dependence. Kaplan {\em et
al.}\cite{Kaplan} showed that the general dependence of such
relaxation rates can have a very complicated dependence on energy
which is unique for each material.  Dynes {\em et al.} however,
used successfully a constant  lifetime broadening  in the
theoretical fits of their measured conductance spectra in low
$T_c$ SIS junctions.\cite{Dynes} Such approximation indeed works
well for fitting of experimental spectra in isotropic
superconductors and in the low transparency limit. In this case,
$\Delta(x)\approx constant$, the conductance is proportional to
the quasi-particle density of states, and this has strong sharp
peaks at $E=\pm\Delta_{0}$. The addition of lifetime broadening as
a small imaginary part to the energy allows one to avoid
divergence at the peak energies and does not produce any essential
change in other regions. Our attempt however, to apply this method
to high-transparency spectra was unsuccessful. We attribute this
to the appearance of additional bound states due to a significant
increase in the decay length (see below).  It is thus clear that
in such cases, taking into account the energy dependence of the
lifetime broadening is essential. In our calculations we used a
fourth degree polynomial approximation for this dependence:

\begin{equation}\label{1}
  \Gamma(E) = \gamma_{1}E^{2}+\gamma_{2}E^{3}+\gamma_{3}E^{4},
\end{equation}

\noindent where the coefficients $\gamma_{i}\leq1$ (in units of
$\Delta_{0}$) were chosen in order to obtain the best fit to the
experimental conductance spectra. Generally, the function $\Gamma$
is added as an imaginary part to the energy in the expression for
the DOS.\cite{Dynes, Tanaka} In our calculations however, the
equivalent of the DOS is the function $\sigma_{S}$ of Eq. (2).
Since we obtain $\sigma_{S}$ by solving two coupled differential
equations in which the energy is a parameter, we could not just
add the $i\Gamma$ term to {\em E} directly. The fact that the
boundary conditions of these equations include nonlinear terms in
{\em E}, complicates things further. We therefore  calculated
$\sigma_{S}$ first without any broadening term, and then added the
broadening numerically by a convolution of the obtained
$\sigma_{S}(E )$ function for each angle $\varphi$, with a
Lorentzian distribution function:

\begin{equation}
L(E) = \frac{1}{\pi}\frac{\Gamma(E)}{(E-E_{0})^{2}+\Gamma^{2}(E)},
\end{equation}
\noindent of width $\Gamma(E)$.\\

We focus now on the case of a pure d-wave symmetry  of the order
parameter without any sub-dominant component. Two basically
different cases of the node ($\alpha = \pi/4$) and anti-node
($\alpha = 0$) orientations will be discussed. $\alpha$ is the
angle between the lobe-direction of the $d_{x^{2}-y^{2}}$-wave
order parameter and the normal to the interface. In the node
direction, a ZBCP exists independently of the values of the
coefficients $k_{1}$ and $k_{2}$ in Eq. (6),  and the effective
barrier strength $Z$. Nevertheless, by varying $k_{1}$ and $k_{2}$
in Eq. (6), the shape of the calculated conductance spectra
changes. Decreasing their values, leads to an increased
suppression of the order parameter at the interface and an
increased decay length. As a consequence, additional
quasi-particle bound-states are created.\cite{Barash, Matsumoto}
These new bound-states are formed in a potential well defined by
the interface on one side and the growing d-wave pair potential
toward the bulk value on the other side. They manifest themselves
as additional peaks in the function $\sigma_{S}(E)$ versus $E$.
Their number and energies vary for different angles $\varphi$, and
according to Matsumoto and Shiba,\cite{Matsumoto} their energies
are given by

\begin{equation}
  E_{n}=\pm
  \Delta_{0}\mid\sin(2\varphi)\mid\sqrt{\frac{n}{r}(2-\frac{n}{r})},
\end{equation}
where $n$ is an integer,
\begin{equation}
   r=\frac{2\mid\sin(2\varphi)\mid}{k_{2}},
\end{equation}
\noindent and $0<n<r$. $k_{2}$ in Eq. (11) is different from zero
since a $k_{2}$=0 in Eq. (6) would yield $\Delta\equiv 0$ which
means that no superconductivity exists. The number of bound states
is determined by $r$. The larger the value of $r$ the larger the
number of these states. Eq. (11) does not give the dependence of
$r$ on $k_{1}$, because in Ref. \cite{Matsumoto} only the special
case of a superconductor-vacuum interface is considered
($Z\rightarrow\infty$). Therefore in Ref. \cite{Matsumoto},
$k_{1}$ is always zero in the node direction. It is however clear
from Eq. (6) that with increasing $k_{1}$ the potential well is
filled up, the spatial dependence of $\Delta$ approaches that of a
step function, and the bound-states disappear. Thus, in the
general case of arbitrary $Z$ and $\alpha$, the number of
bound-states is proportional to the depth of the potential well.
This depends on both coefficients $k_{1}$ and $k_{2}$, and yields
a more general expression for $r$ as follows:

\begin{equation}
  r=\frac{2\mid\sin(2\varphi)\mid}{k_{2}}[1-\tanh(k_{1}k_{2})].
\end{equation}

Eq. (12) shows that while we consider quasi-particles within a
narrow cone around $\varphi=0$ (like in SIN junctions),
bound-states will play a less significant role in the conductance
(if the cone angle $\varphi\rightarrow 0$ then there are no bound
states since $r\rightarrow 0$). In the weak barrier limit though,
a broad cone of about $-\pi/2<\varphi<\pi/2$ should be used, and
the influence of bound-states on the conductance spectra becomes
substantial as seen in Fig. 2. On each curve of this figure only
two symmetrical maxima besides the ZBCP are present. The curves in
Fig. 2 however, show the $\sigma(eV)$ dependence on V and not the
$\sigma_{S}(E)$ dependence on E. In the later,  for each value of
$\varphi$  a different number of additional maxima exist. After
integration of $\sigma_{S}$ over all angles $\varphi$, only two
main peaks versus energy are left which contribute to the
resulting maxima $E_{pd}$ in $\sigma(eV)$ as seen in Fig. 2. These
side band peaks are the second most prominent features in the
spectra. With increasing $r$, the voltage of these peaks shifts to
low bias. Because of this, after the integration in Eq. (2) the
ZBCP becomes somewhat smaller and its width decreases. It is
important to note, that in spite of the shift of additional maxima
towards zero bias with increasing decay length of the pair
potential, they can't annihilate the ZBCP, since the zero-energy
bound state is robust and exists for {\em all} values of
$\varphi$, while the other bound-states are not. Therefore, by
changing the parameters $k_{1}$ and $k_{2}$ we can't obtain
splitting of the ZBCP in the node direction in the pure
d-wave case.\\

Splitting of the ZBCP along the node direction was however
measured in SNS junctions.\cite{Koren} In order to fit this data,
it was necessary to add a sub-dominant component to order
parameter with a different symmetry. Because of previous
results,\cite{Covington,Krupke} we have choosen for this purpose
to add an imaginary s-wave component to the pair potential. Now
the form of the resulting order parameter depends on all four
coefficient $k_{1},k_{2},k_{3}$ and $k_{4}$. Our calculation
results along the node direction and with a small $Z$ value are
shown in Fig. 3. The main conclusion from this figure is that the
sub-dominant component can not essentially influence the
conductance spectra as long as the decay length of the dominant
d-wave component is $\sim \xi_{0}$. This behavior is different
from that observed in the strong barrier limit, where a small
decay length for both components is enough for obtaining splitting
of the ZBCP (not shown). Tanuma {\em et al.} have obtained no
sub-dominant component for low values of $Z$ ($Z\sim
1$),\cite{Tanuma} even though their $Z$ value is defined as twice
the BTK $Z$ value that we use here. In addition, the order
parameter that they obtained, almost always had a decay length of
order $\sim\xi_{0}$. However, when we repeated their self
consistent calculations for high transparency junctions, we could
find cases (when $T_{cs}=0.5T_{cd}$ for instance) in which the
s-component did exist and had a long penetration length of the
order of 10$\xi_0$ into the bulk. Moreover, in their
self-consistent calculations of the order parameter they did not
take into account the proximity effect. Therefore, the absence of
a sub-dominant component in the results of Tanuma {\em et al.} for
weak barriers, is not contradictory to our results. As can be seen
in Fig. 3,  we obtain the splitting effect by the s-wave component
only when d-wave decay lengths longer than a few $\xi_0$ are
used.\\

A sub-dominant $is$ component splits the ZBCP into two peaks,
whose energy $|E_{ps}|$ depends on the values of $k_{3}$ and
$k_{4}$ in Eq. (7). There is no exact analytic formula for this
dependence, but numerical calculations yield an approximate
relation:
\begin{equation}\label{}
  |E_{ps}| \sim c_{1}k_{3}\exp(-c_{2}k_{4})\,\,\,\,\,\,\,\,\,
  \,\,\,\,for\,\,\,\, k_{3}<\Delta_{0},
\end{equation}
\noindent where  $c_{1}$ and $c_{2}$ are constants which depend on
$Z$. When two components of the order parameter coexist, the
conductance spectra is a superposition of contributions containing
the bound-states of both. These bound-states can annihilate one
another unless the value of $E_{ps}$ is close to the energy
$E_{pd}$ of the bound-states of the dominant component. It thus
follows that the smaller the s-wave component, the larger the
required decay length of the d-component, which is needed in order
to obtain a clear
splitting.\\

A simulation of a the conductance spectra  fitting our previous
experimental data along the node direction is given in Fig. 4. The
original data was obtained for SNS junctions.\cite{Koren} Since no
theoretical model is available at the present time for the
calculation of the  conductance of SNS junctions which takes into
account the finite decay length of the pair potential at the
interface, we used as a first approximation the SN model discussed
here, but with a re-normalized energy scale to fit the SNS data.
This re-normalization was simply done by doubling the energy scale
in the SN calculations (multiplying it by a factor of 2). We can
thus compare the modified SN simulation results with the measured
SNS data on the same energy scale of $E/\Delta_0$. Coexistence of
both $d$ and $is$ components of the order parameter near the
interface had to be assumed in order to fit the data. The good
agreement between theory and experiment indicates that large decay
lengths of the d-wave pair potential are actually present
near the interface of real junctions.\\

We now turn to junctions along the anti-node direction. This
direction is characterized in the tunneling regime by a "V" shaped
conductance spectra with a minimum at zero bias whose depth is a
function of the barrier strength $Z$. For a low transparency
barrier $\sigma(E=0)\sim 0$, but as $Z$ decreases this value
increases.  For weak barriers (small $Z$) the conductance is
controlled mostly by Andreev reflections,  and $\sigma(E=0)$
reaches a value of 2 when $b=0$ and $a=1$ in Eq. (4). Again, we
varied the values of the coefficients $k_{i}$ ($i=1..4$) in order
to study the influence of the order parameter near the interface
on the conductance spectra. Qualitatively, all our previous
reasoning remain valid. Decreasing $k_{1}$ and $k_{2}$ leads to
the appearance of bound-states which are seen as new maxima in the
simulated conductance curves. Similar to the results along the
node direction, we find that with increasing decay length of the
d-wave pair potential, the bound-states appear at lower bias. But
they never occur at zero-bias, where the conductance minimum
persists up to very low $Z$ values. Fig. 5 shows simulations of
conductance spectra for a weak barrier of $Z=0.5$ along the (100)
direction for different values of the coefficient $k_{2}$ while
$k_{1}=0$. The main feature of these spectra is that the peaks at
the sub-gap energies increase with decreasing $k_{2}$ (increasing
decay length). Moreover, addition of a small sub-dominant s-wave
component to the d-wave order parameter when a large decay length
is used, does not practically change the form of the conductance
spectra (The solid and dotted  curves in Fig. 5 are almost
indistinguishable). This leads to an important conclusion that the
conductance along the (100) direction in high transparency
junction with a long-range proximity effect, is basically
insensitive to the presence of
possible sub-dominant components of order parameter.\\

Fig. 6 shows experimental conductance spectra of a
low-transparency junction along the (100) direction,\cite{Koren}
together with a best fit simulation for this case. The energy
scale of the SIN simulation was re-normalized to the equivalent
SIS scale of the experiment as before (was multiplied by a factor
of  2). The low transparency is a result of a large normal
resistance $R_N$ which is found in the experiment for the (100)
orientation. The roughness of the interface in this type of
junction was found to be smaller than the coherence length
$\xi_{0}$, thus no faceting can lead to the low transparency.
Twinning however, can cause different $R_N$ values in different
directions. We found out that a pure d-wave order parameter
neither with a small nor with large decay length can fit our data
well, especially not at low bias (see the dash and dotted curves
in Fig. 6). It turned out that the contribution of bound states is
essential in order to describe correctly  the low energy
dependence of the experimental conductance spectra. A typical
contribution of bound states is shown in Fig. 6 by the dotted
curve for a $20\xi_0$ decay length. For the best fit in this
figure we used a superposition of mainly a constant d-wave
component (76\%), and the additional contributions from three
longer decay lengths of $3\xi_0$, $10\xi_0$ and $20\xi_0$ (8\% of
each), and found a remarkable agreement between the simulation and
the experimental data (the solid line in Fig. 6).  The fact that
we had to use such a superposition means that in reality, at least
in the lobe direction, there is some non-uniformity in the values
of the decay length over the junction cross section. This can be
due to the twinning mentioned above, to a non-uniform oxygen
distribution or to some other disturbances. We note that in grain
boundary junctions a nonuniform distribution of different barrier
width was found,\cite{Mannhart} which in a way is similar to the
decay
length distribution observed here.\\

In the following we summarize the main properties of  SN and SIN
junctions resulting from our simulations:\\
 i)For a pure d-wave superconductor in the node-direction,  large decay lengths of the
 order parameter near the interface lead to additional peaks in
 the conductance spectra at sub-gap biases. The peak energies shift to
 lower bias with increasing decay length of the order
 parameter, while the ZBCP amplitude and its width decrease.\\
 ii)In order to obtain a well defined splitting of ZBCP
 in the conductance spectra along the node direction as seen in the experiment,
 one has to add a small sub-dominant $is$
 component to the order parameter, and keep a long decay length.\\
  iii)In contrast, using a pure d-wave with a large decay length in
  high transparency junctions along the lobe
  direction, the gap peaks in the conductance move to lower energies
  and this looks like splitting of the broad Andreev-enhanced conductance
  regime at $E\lesssim \Delta_0$.\\
 iv)The energy dependence of the quasi-particles lifetime broadening has to
 be taken into account in the simulations of the conductance
 spectra.\\
 v)The fact that a long decay length in the superconductor had to be used
 in order to obtain a good agreement between the theoretical simulations and
 the measured conductance spectra, gives further support to the notion that
 a long-range proximity effect exists in the cuprates.\\

 \maketitle{\large CONCLUSIONS}

In the present study we investigated how the spatial dependence of
the order parameter near the interface of junctions affects the
resulting conductance spectra. We found that the decay length of
the d-wave pair potential is the most significant parameter
affecting the spectra. More specifically, for high transparency
junctions along the node direction, the use of a long decay length
of the order of $10-20\xi_0$ yields a splitting of the ZBCP,
provided an additional small $is$ component exists near the
interface. If in contrast a $\sim\xi_0$ decay length is used, no
such splitting develops in the simulations, and this is in
contradiction to experimental observations. For low transparency
junction along the main lobe direction, a small decay length of
the order of $\sim \xi_0$ could not describe the data neither with
a pure d-wave nor with a d+is order parameter. Only when a
distribution of decay lengths between $0\xi_0$ and $20\xi_0$ was
assumed, contributions from additional bound states led to a good
fit of the experimental data. Another important parameter that
affects the conductance spectra strongly is the quasi-particles
lifetime broadening and its non-trivial energy dependence
($\Gamma(E)$). We actually modified its basic Fermi liquid $E^2$
dependence to fit the experimental data along the node direction,
and then used it successfully without any further changes to fit
the lobe data. Finally, we note that a long decay length in the S
side is consistent with a reversed proximity effect in high
transparency SN junctions, and the low $T_c$ of the $is$
component.\\

{\em Acknowledgments:}  This research was supported in part by the
Israel Science Foundation, the Heinrich Hertz Minerva Center for
HTSC, the Karl Stoll Chair in advanced materials, the Fund for the
Promotion of Research at the Technion, and by the Center for
Absorption in Science, The Ministry of Immigrant Absorption of the
State of Israel.\\

\newpage

{\bf \large Figure Captions}

\begin{description}

\item [Fig. 1:] Typical spatial dependencies of the
d-wave and s-wave pair potentials near the interface, used in our
simulations. The pair potential axis is normalize to $\Delta_{0}$,
the bulk value of the superconductor's gap.

\item [Fig. 2:] (color) Simulations of normalized
conductance spectra of a high transparency SN junction with Z=0.5
along the node direction for a pure d-wave symmetry of the pair
potential. Different curves correspond to different values of the
the decay length.

\item [Fig. 3:](color). Theoretical curves of normalized conductance spectra of a
high-transparency NS junction along the node direction using a
d+is symmetry of the pair potential. In each curve the decay
length of both components of the pair potential is the same,
except for the dash double dotted line where  the decay length of
the d-wave component is $\xi_0$ while that of the s-wave component
is 18$\xi_0$.

\item [Fig. 4:] (color) Measured normalized
conductance spectra of a high transparency
$YBa_{2}Cu_{3}O_{6+x}/YBa_{2}Fe_{0.45}Cu_{2.55}O_{6+x}/YBa_{2}Cu_{3}O_{6+x}$
junction along the node direction (symbols) together with a best
fit simulation (line). The broadening $\Gamma$ versus E is given
in the inset.

\item [Fig. 5:] (color) Calculated normalized conductance spectra of a high-transparency
SN junctions with Z=0.5 along the lobe direction. All curves were
obtained by the use of a pure d-wave symmetry of the pair
potential with various decay length, except for the dotted line
where a d+is symmetry was used.

\item [Fig. 6:] (color) Measured normalized conductance spectra along the lobe direction
(symbols) of a low transparency junction of\\
$YBa_{2}Cu_{3}O_{6+x}/YBa_{2}Fe_{0.45}Cu_{2.55}O_{6+x}/YBa_{2}Cu_{3}O_{6+x}$
together with a best theoretical fit (solid line). The energy
dependent lifetime broadening $\Gamma(E)$ is the same as that in
Fig. 4. Also shown are the conductance contributions of a constant
d-wave (dash line) and a d-wave with 20$\xi_0$ decay length
(dotted line).

\end{description}

\newpage

\bibliography{AndDepBib.bib}

\bibliography{apssamp}% Produces the bibliography via BibTeX.

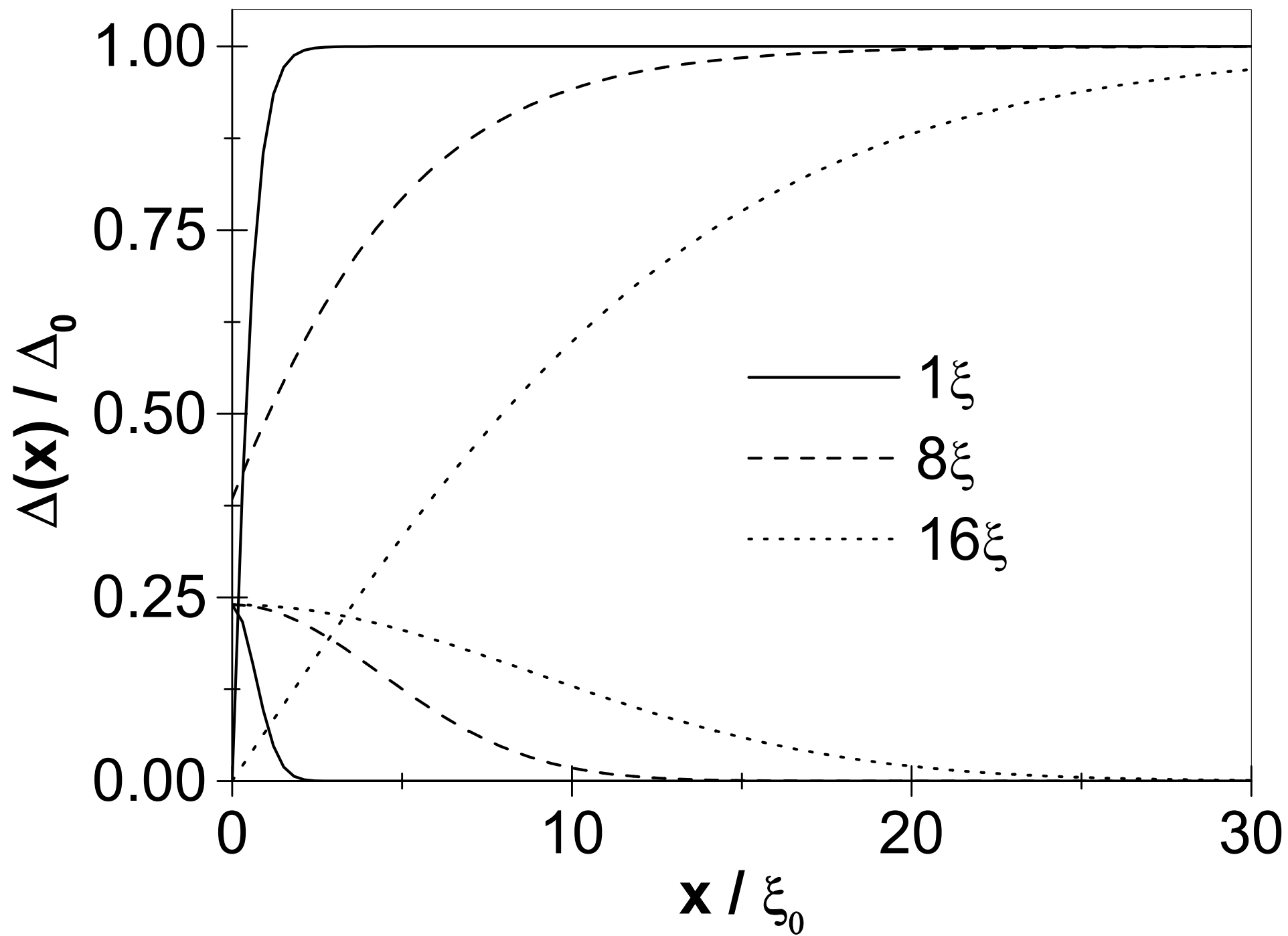
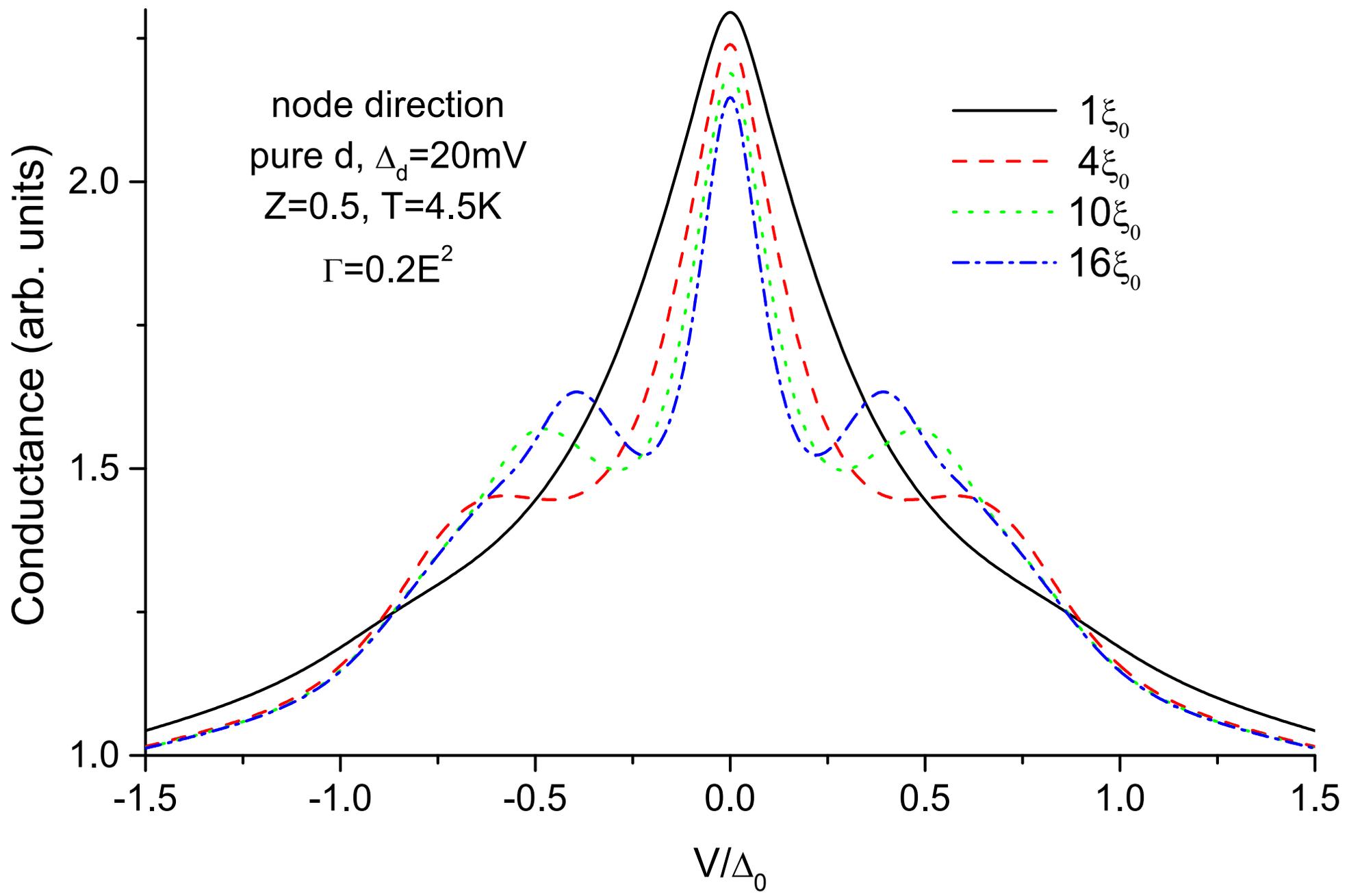
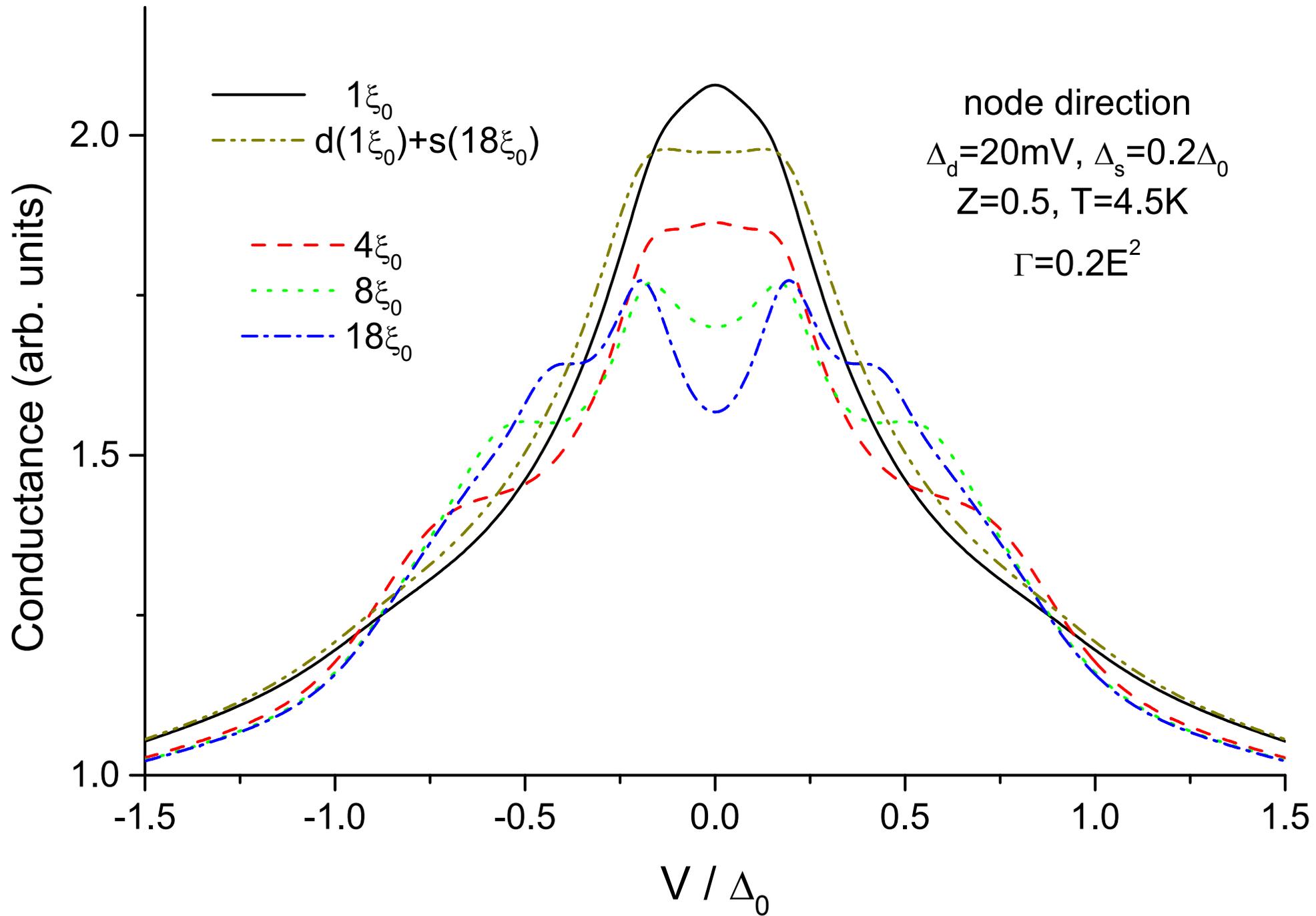
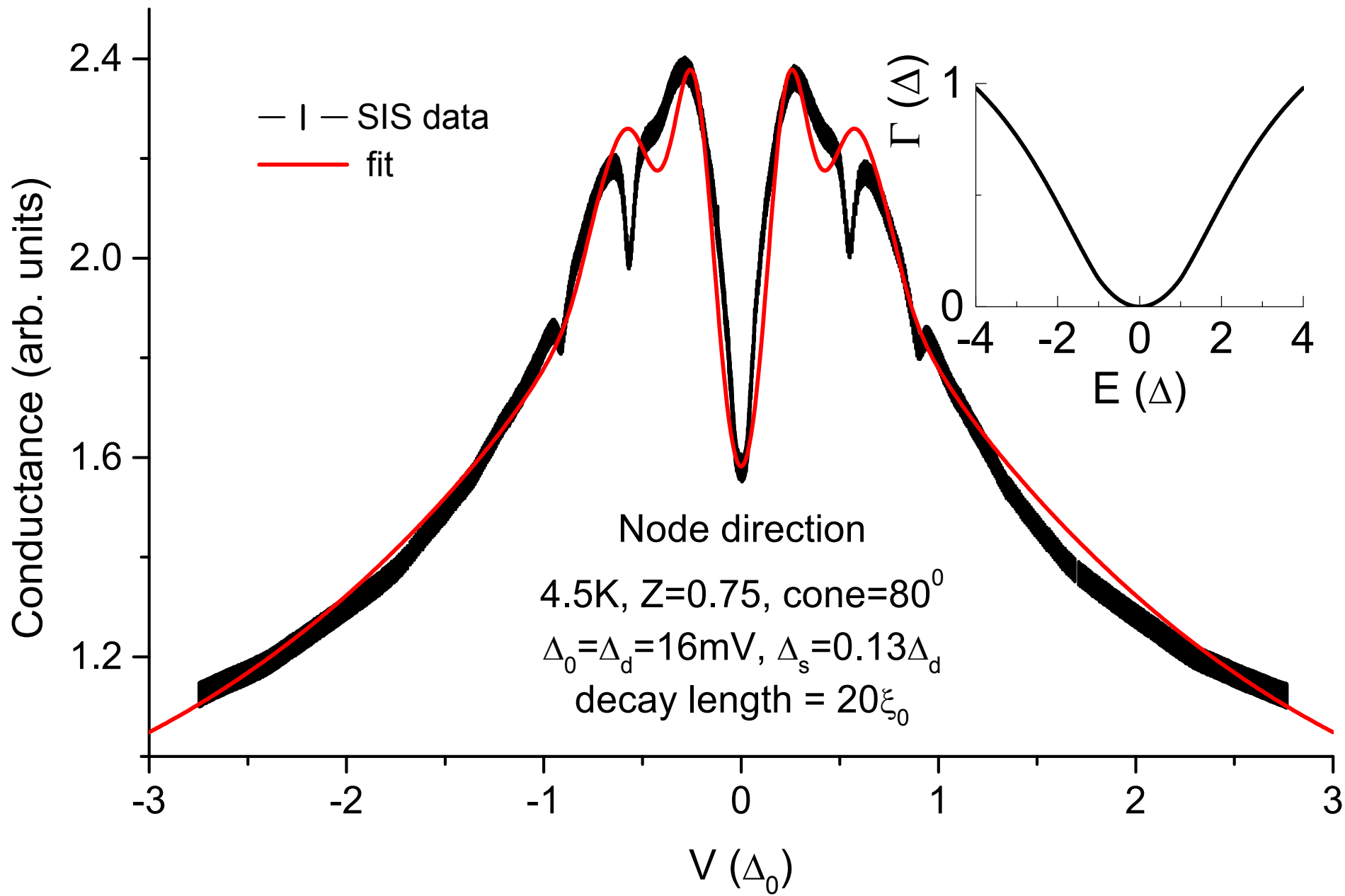
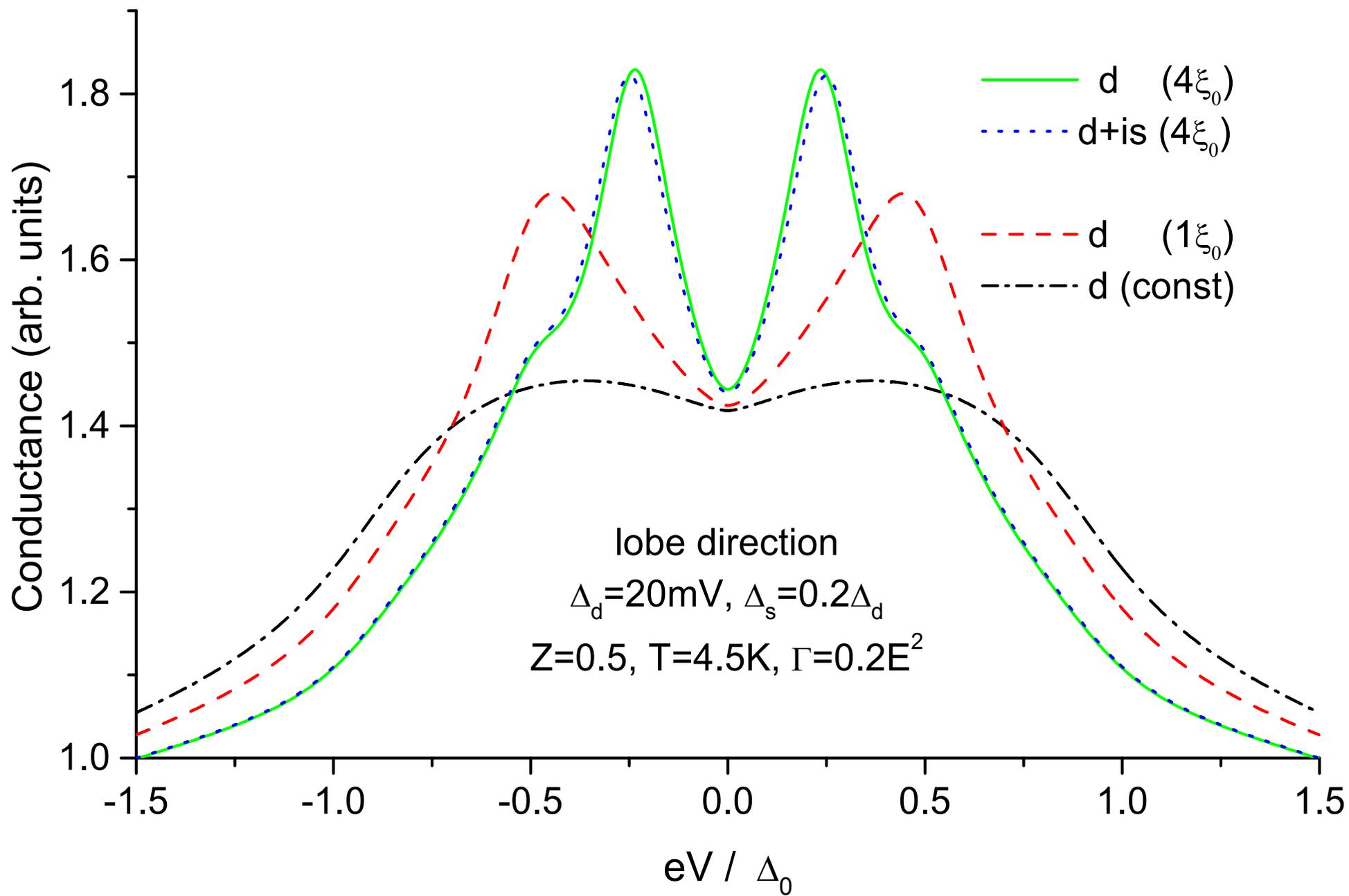
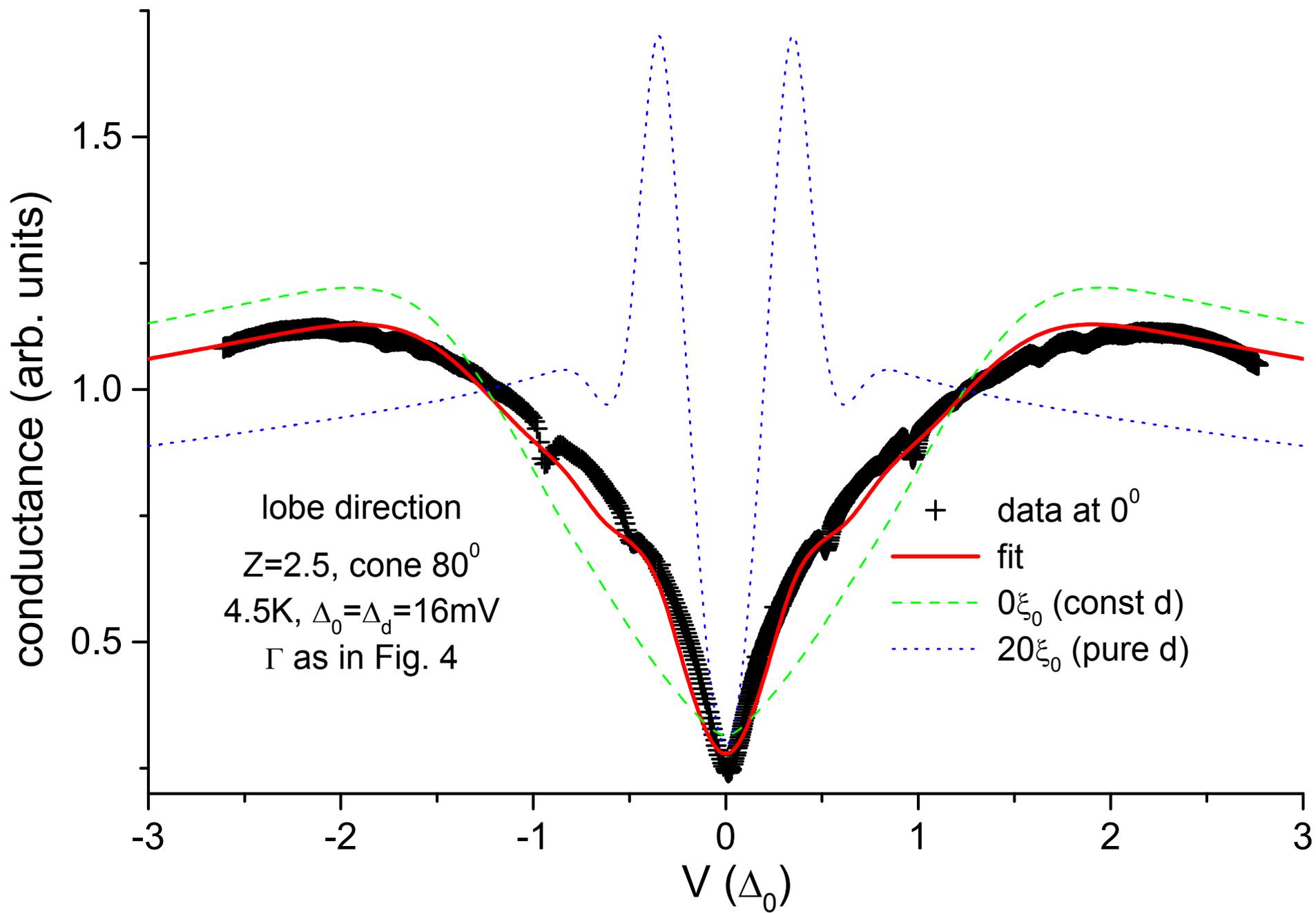
\end{document}